# Investigating the impact of COVID-19 on Online Learning-based Web Behavior

Nirmalya Thakur[1], Saumick Pradhan[1], Chia Y. Han[1]

[1] Department of Electrical Engineering and Computer Science, University of Cincinnati, Cincinnati, OH 45221-0030, U.S.A.

## ABSTRACT

COVID-19, a pandemic that the world has not seen in decades, has resulted in presenting a multitude of unprecedented challenges for student learning across the globe. The global surge in COVID-19 cases resulted in several schools, colleges, and universities closing in 2020 in almost all parts of the world and switching to online or remote learning, which has impacted student learning in different ways. This has resulted in both educators and students spending more time on the internet than ever before, which may be broadly summarized as both these groups investigating, learning, and familiarizing themselves with information, tools, applications, and frameworks to adapt to online learning. This paper takes an explorative approach to further investigate and analyze the impact of COVID-19 on such web behavior data related to online learning to interpret the associated interests, challenges, and needs. The study specifically focused on investigating Google Search-based web behavior data as Google is the most popular search engine globally. The impact of COVID-19 related to online learning-based web behavior on Google was studied for the top 20 worst affected countries in terms of the total number of COVID-19 cases, and the findings have been published as an open-access dataset. Furthermore, to interpret the trends in web behavior data related to online learning, the paper discusses a case study in terms of the impact of COVID-19 on the education system of one of these countries.

# INTRODUCTION

Towards the end of 2019, the Huanan Seafood Wholesale Market in Wuhan, Hubei, China, experienced an outbreak of strange pneumonia accompanied by fever, dry cough, and fatigue, which was also accompanied by gastrointestinal problems (Huang et al. 2020). The outbreak involved about 66% of the staff in the market, which resulted in the same getting closed on January 1, 2020, to prevent the spread of the virus. However, the virus had already started to spread by then both in China and in different countries of the world via human-human transmission, community spread, surface-based transmission, and community spread (Català et al. 2020, Chakraborty et al. 2020). This virus was soon declared as a pan-demic by the World Health Organization (WHO) on account of the associated ill-ness, hospitalizations, and deaths that it was causing on a global scale (WHO 2020). This virus was named severe acute respiratory syndrome coronavirus 2 (SARS-CoV-2) or COVID-19 by the International Committee on Taxonomy of Viruses (Gorbalenya et al.). At the time of writing this paper, there have been more than 243 million cases of COVID-19 and over 4.9 million deaths on a global scale (WHO 2021, Worldometers 2021).

To reduce the spread of the virus, schools, colleges, and universities in almost all the countries started switching to online or remote learning in 2020 (Dhawan 2020, Daniel 2020). Online learning may broadly be defined as the use of the internet and associated tools (devices, gadgets, platforms, software, etc.) for instructional delivery and management. Online learning can be synchronous or asynchronous (Anderson 2008), and the success and effectiveness of online learning depend on the preparation, familiarity, and seamless use of these associated tools by both educators and learners during the asynchronous or synchronous delivery of instruction (Hrastinski 2008). Therefore, the transition to online learning was associated with both educators and students spending more time on the internet than ever before, which may be broadly summarized as both these groups investigating, learning, and familiarizing themselves with information, tools, applications, and frameworks to adapt to online or remote learning. Internet activity is characterized by web behavior (Soler-Costa et al. 2021) and can be further studied to interpret the associated interests, challenges, and needs demonstrated by the specific kind of web-behavior. As Google is the most popular search engine on a global scale (Statista 2021), therefore we focused on studying the web behavior related to online learning from Google for this work.

Previous works in this field in the context of online learning during COVID-19 have focused on evaluating student perceptions (Bączek et al. 2021), studying barriers to online learning (Baticulon et al. 2021), investigating the effectiveness of learning strategies (Tsang et al. 2021), application of self-determination theory to explain student engagement (Chiu et al. 2021), investigating readiness for online learning (Rafique et al. 2021), factors to motivate learning (Chiu et al. 2021), studying attitudes of students (Hussein et al. 2020), investigating student learning outcomes and satisfaction (Baber 2021), exploring strategies to increase student interest (Sutarto et al. 2020), studying adaptation towards online learning (Xhelili et al. 2021), degrees of parental engagement (Novianti et al. 2020), and effectiveness of online learning through WhatsApp groups (Susilawati et al. 2020). However, none of these works focused on studying the patterns of interest towards online learning emerging from different countries of the world due to COVID-19. Studying the same at a country-level would help to interpret the associated effectiveness, challenges, and needs related to wide-scale adoption and implementation of online learning in the education sector of each of the countries involved. Therefore, addressing this research challenge by studying the relevant web behavior data in terms of search interests originating from the top 20 worst-affected countries due to COVID-19 serves as the main motivation for this work.

## METHODOLOGY AND RESULTS

This work was performed using Google Trends (Google 2021). Google Trends is a website developed by Google that allows tracking web behavior patterns related to different search terms on Google originating from different parts of the world. In contrast to traditional surveys, Google Trends has several advantages, as outlined in (Mellon 2013). In view of these advantages of Google Trends, it was selected for this study. Google Trends uses a scale from 0 to 100 to represent the popularity related to a search term on Google in terms of search interests. This numerical value represents a weighted result based on the topic's proportion to all searches on all topics recorded on Google. The number from 0 to 100 is determined by dividing the data point by the total searches of the area and the time period that it represents so that the algorithm is able to compare the relative popularity of the search term and assign it an appropriate number in this scale. For this study, the search term that was selected was "online learning" and the corresponding data for the 20 worst affected countries in terms of COVID-19 (at the time of writing of this paper) - U.S.A., India, Brazil, U.K., Russia, France, Turkey, Iran, Argentina, Colombia, Spain, Italy, Indonesia, Germany, Mexico, Poland, South Africa, Philippines, Ukraine, and Peru, were mined using Google Trends. Table 1 represents the number of COVID-19 cases and the associated deaths in each of these countries so far (Worldometer 2021).

**Table 1:** The top 20 worst affected countries in terms of COVID-19 cases

| Country Name | COVID-19 cases | No of deaths |
|---|---|---|
| USA | 44,983,901 | 728,884 |
| India | 33,914,012 | 450,152 |
| Brazil | 21,517,514 | 599,414 |
| UK | 8,046,390 | 137,417 |
| Russia | 7,690,110 | 213,549 |
| France | 7,043,316 | 116,991 |
| Turkey | 7,357,336 | 65,590 |
| Iran | 5,674,083 | 122,012 |
| Argentina | 5,263,219 | 115,379 |
| Colombia | 4,965,847 | 126,487 |
| Spain | 4,971,310 | 86,701 |
| Italy | 4,692,274 | 131,198 |
| Indonesia | 4,224,487 | 142,494 |
| Germany | 4,300,536 | 94,861 |
| Mexico | 3,699,621 | 280,607 |
| Poland | 2,916,969 | 75,803 |
| South Africa | 2,908,768 | 87,981 |
| Philippines | 2,632,881 | 38,937 |
| Ukraine | 2,497,643 | 57,840 |
| Peru | 2,181,183 | 199,559 |

For each of these countries, we mined the data related to "online learning" on a monthly basis from Google Trends in terms of the associated search interests on a scale of 0 to 100. The earliest month for which this data was available was January 2004, and the most recent data was available for October 2021. This data from Google Trends and the data shown in Table 1 were collected on October 7, 2021. Upon data collection, we calculated the mean value of search interests related to "online learning" by considering two timeframes – "before COVID-19" and "after COVID-19". To decide on the distinction between these timeframes, we considered the declaration of COVID-19 as a pandemic by WHO to be the distinction point in the timeline. Or in other words, January 2004 to March 2020 was considered to be "before COVID-19," and April 2020 to October 2021 was considered to be "after COVID-19". The mean value of search interests for each of these timeframes was calculated by taking into consideration the monthly search interest values in each time period. Table 2 shows the results of the same. We would wish to state a couple of considerations for this study. First, the data related to the COVID-19 cases, as shown in Table 1, is based on the data available in (Worldometer 2021). There are various other online sources that list the number of COVID-19 cases in a specific geographic region. There can be various reasons, as mentioned in (Koch et al. 2020), which can cause the reporting of COVID-19 cases in a specific geographic region to slightly differ across different online platforms or tools. To the best of our knowledge, this source (Worldometer 2021) is a reliable source for tracking the number of COVID-19 cases in different geographic regions. Therefore, the data from the same was used for this study. The search interest-related data from Google was collected from January 2004 to October 2021. However, the study was conducted on October 7, 2021, so the data for October 2021 is based on 7 days. This data for October 2021 can change at a later point in time based on the web behavior patterns recorded in the remaining days in the month of October.

**Table 2:** Comparison of Search Interests related to Online Learning for all the countries before and after COVID-19

| Country Name | Search Interest related to Online Learning | |
|---|---|---|
| | **Before COVID-19** | **After COVID-19** |
| USA | 49.96907 | 50.20 |
| India | 29.96907 | 20.30 |
| Brazil | 8.134021 | 9.000 |
| UK | 39.05670 | 54.25 |
| Russia | 3.788889 | 1.500 |
| France | 10.21649 | 10.00 |
| Turkey | 5.814433 | 6.050 |
| Iran | 7.626316 | 1.470 |
| Argentina | 3.386598 | 5.350 |
| Colombia | 5.922680 | 10.80 |
| Spain | 24.90206 | 36.15 |
| Italy | 16.17010 | 26.90 |
| Indonesia | 4.639175 | 7.550 |
| Germany | 39.75258 | 50.15 |
| Mexico | 9.592784 | 23.65 |
| Poland | 15.30928 | 20.50 |
| South Africa | 13.60309 | 37.75 |
| Philippines | 7.881443 | 60.25 |
| Ukraine | 4.829897 | 7.150 |
| Peru | 3.417526 | 5.950 |

## DATASET DESCRIPTION

The data for each of these months for each of these countries was compiled and published as an open-access dataset available at https://dx.doi.org/10.21227/pa7d-nt11. The dataset consists of one .csv file with the file name - "Online_Learning_Data.csv". This file contains 21 data attributes. The first attribute, "Month," stands for the month from January 2004 to October 2021, as the data was collected on a monthly basis in this range. The remaining 20 attributes represent the search interests related to "online learning" for each of the 20 countries that have been worst affected on account of COVID-19, as shown in Table 1. This data was directly compiled from Google Trends, so the minimum value is 0, and the maxi-mum value is 100 for the search interests. As per Google Trends, the minimum value of 0 represents minimal search interest or no interest in the search topic, and the maximum value of 100 represents maximum search interest in the search topic. For certain countries, for a few months, almost no internet activity was recorded for the search terms "online learning", so Google Trends did not provide a numerical value. This is equivalent to the condition of the search interest value being 0; therefore, we assigned a numerical value of 0 for all these months for each of the countries where such a situation was recorded. This dataset has been published as open access to further support research and development in this field.

## CASE STUDY

In this section, we outline the case study that we performed for one of these countries, India, to interpret the change in the numerical value of search interests related to "online learning" in the two timeframes – "Before COVID-19" and "After COVID-19". As shown in Table 2, it can be observed that the search interest related to "online learning" significantly decreased in the "After COVID-19" timeframe for India. So, we studied and analyzed the factors, circumstances, incidents, and events related to the education system in India and its response to COVID-19 to interpret this change. India is the second most populous (1.353 billion people), seventh-largest country by area, and has the second-largest number of students in the world (Pre-primary 71,285,198, Primary 123,996,484, Secondary 177,432,640, Tertiary 122,369,632, by education level) (British Council 2019). Its education system is one of the oldest in the world and was introduced based on traditional fundamental concepts associated with different Indian religions. It is now governed by both federal and state governments (Rajashree et al. 2014, Sharma et al. 1996). Before the COVID-19 pandemic, India was following its National Policy on Education from 1986 (Roy 2020). Due to the onset of the COVID-19 pandemic, the Indian government started a nationwide lockdown for all educational institutions on March 16th, 2020 (Khattar 2020), which caused a rapid transition to online learning across all schools, colleges, and universities across India.

This rapid transition caused several issues for senior faculty members (especially from the rural areas) as they were not familiar with online platforms and tools such as Zoom, WebEx, Teams, etc. This nationwide lockdown resulted in students across India losing nearly 3 months of their academic year. The transition to online learning was difficult for students in government schools, especially from rural areas. In January 2020, India had only 560 million total active users of the internet, which is about half of its population (MoSPI 2021). To add, approximately 50% of the households in India have slow internet connections (IGUAGE 2020), and 27% of students do not have access to a smartphone or computer (Kapasia et al. 2020). These challenges caused decreased participation and "interest" of students across India towards online learning. The transition and the associated multimodal challenges to online learning also affected the mental health of the students (Lathabhavan et al. 2020). The unemployment rate shot up from 8.4% in mid-March to 23% in early April, and the urban unemployment rate increased to 30.9% (Jena 2020). This caused students from poor financial backgrounds to start working instead of participating or adapting to online learning.

Furthermore, several students from poor financial backgrounds used to depend on mid-day meals (a kind of meal provided to students in certain government schools in India) for their sustenance (Singh et al. 2014). With the onset of the pandemic and the associated lockdown, such students were strained for resources to have proper meals during the day. Furthermore, the diversity across the dense population of India in terms of cultural background, language barriers, communication styles, learning approaches, and living styles created multiple hurdles towards the adoption of online learning in one specific language, such as English. To summarize, through this comprehensive case study, we highlighted the factors, circumstances, incidents, and events related to the education system in India and its response to COVID-19. The study outlines some of the main reasons which point towards the fact that the rapid transition to online learning was not a success in different parts of India, which is further supported by the decrease in the search interest value related to "online learning," as shown in Table 2. The study also discusses potential areas of improvement and areas where remedial measures can be implemented by policymakers to improve the adoption of online learning across India. Similar case studies can be performed for all the other 19 countries mentioned in Table 2 to provide information to policymakers and governments to support and enhance the adoption of online learning on a global scale during the COVID-19 pandemic. For paucity of space, the case study for only one country has been included in this paper.

## CONCLUSION AND FUTURE WORK

The declaration of COVID-19 as a pandemic by the WHO was followed by schools, schools, colleges, and universities closing in 2020 in almost all parts of the world and switching to online or remote learning, which has impacted student learning in different ways. This rapid transition towards online learning has been accompanied by both educators and students spending more time on the internet to familiarize themselves with various platforms and tools, the knowledge of which is necessary to facilitate and sustain online learning-based courses. Such internet activity, recorded in the form of web behavior, can be mined and studied to interpret and investigate the associated search interests towards online learning, which would provide grounds for the analysis of the associated challenges and needs related to online learning to evaluate the degree of success towards this transition. To investigate this research challenge, in this paper, we studied the web behavior related to online learning emerging from the 20 worst affected countries in terms of the number of COVID-19 cases, by using Google Trends. The search interest data, on a monthly basis, for all these 20 countries was compiled and published in the form of an open-access dataset available at https://dx.doi.org/10.21227/pa7d-nt11. Thereafter, we compared the associated search interest values related to online learning both before and after COVID-19 for each of these countries to interpret the associated trends. Furthermore, we performed a comprehensive case study for one of these countries, India, to interpret the trends in search interests related to online learning on account of COVID-19 with an aim to highlight the factors, circumstances, incidents, and events related to the education system in India and its response to COVID-19. Future work would involve extending this study to include all the countries affected by COVID-19.

## AUTHOR CONTRIBUTIONS

Conceptualization, N.T..; Methodology, N.T. and S.P.; Data Curation, S.P.; Formal Analysis, S.P. and N.T.; Data Visualization and Interpretation, N.T. and S.P..; Results, S.P. and N.T.; Writing-Original Draft Preparation, N.T.; Writing-Review and Editing, N.T; Supervision, N.T; Project administration, N.T and C.Y.H.; Funding Acquisition, Not Applicable. All authors have read and agreed to the published version of the manuscript.

Igauge.in. Available at:
https://www.igauge.in/admin/uploaded/report/files/QSIGAUGECOVIDISPReportA
pril2020_1606732097.pdf (Accessed: January 18, 2022).